\documentclass[9pt, a4paper, twocolumn]{article}

\makeatletter
\let\ps@plain\ps@empty
\ps@empty
\makeatother

\usepackage{amsfonts}
\usepackage{graphicx}
\begin{document}
\sf
\title{\normalsize\textbf{THE MOLECULAR DIFFUSION OF ICE CRYSTALS OF VARIOUS SHAPES}
\newline}
\author{\normalsize Hyun Youk, Roland List, and Theophilus Ola\\
\small.\\
\normalsize Department of Physics, University of Toronto, Toronto, Ontario, M5S 1A7, Canada \\
\normalsize (March 16, 2004)} 
\date{}
\maketitle

\normalsize
\begin{flushleft}
\textbf{1. INTRODUCTION}
\end{flushleft}

 Two main controlling factors of ice crystal growth are the heat and (gaseous) mass transfers, H\&MT, as characterized by the Nusselt number (Nu) for heat and the Sherwood (Sh) number for mass.  Nu and Sh express the increase of the molecular diffusions by the relative motions of the particles in air as characterized by the Reynolds number (Re). Traditionally Nu is assumed to be identical to Sh.  Past laboratory measurements with various ice crystal shapes by the Toronto Cloud Physics Group involved Sh, thus this study will deal with Sh.  The growth of a crystal depends on its size and shape, its fall speed (through Re) and the flux of water vapour from the surrounding air.  It is controlled by the diffusion of heat which carries away the energy released on the crystal surface by deposition. Sh is a function of Re and was established by Schemenauer and List (1978)[S\&L78] for snow crystals and graupel.  Their study, however, did not address the case of pure diffusion (i.e. Re=0).  Such values based on approximations have been available in the literature (McDonald, 1963[Mc63], Jayaweera (1971)) only for a very limited number of shapes of crystals. Thus, it was decided to take a general approach to numerically calculate $Sh_{Re=0}$ for any ice crystal with a rectilinear shape.  New values of $Sh_{Re=0}$ for additional shapes of crystals are addressed in this study along with the method developed for the computations.

 The steady state diffusion is controlled by the Laplace equation.  It was solved numerically with Dirichlet boundary conditions on a rectilinear 3 dimensional lattice system with variable lattice separation distances for hexagonal plates, hexagonal cylinders, stellar crystals, capped columns, and broad-branched crystals.

\hfill

\begin{flushleft}
\textbf{2. ELECTROSTATIC ANALOGY}
\end{flushleft}

 The water vapour density field ($\rho$) around a stationary ice crystal growing under steady state diffusion obeys the Laplace equation
\begin{equation}
\label{1}
\nabla^{2}\rho = 0
\end{equation}

 The initial growth of an ice crystal by the net deposition of water vapour in steady state diffusion, is 
controlled by the rate at which vapour diffuses to and from the crystal's surface.  The rate at which heat is released by the crystal is exactly the same as the rate at which heat is diffused from the surface.  This balance leads to quasi constant value of crystal's surface temperature, and this in turn leads to a constant local vapour density $\rho_{0}$ around the crystal.  For a perfect capacitor that is geometrically similar to the crystal, the electrostatic potential \emph{V} in vacuum also obeys the Laplace equation, with the surface electrostatic potential $V_{0}$ constant as is the case for perfect capacitors.  Identifying \emph{V} with $\rho$, the direct analogy between steady state diffusion and electrostatics is established.  A capacitor that is geometrically similar to the crystal is then set in a rectangular box, a Faraday cage, sufficiently afar from its edges.  The capacitor is then surrounded by the electrostatic potential obeying Laplace equation.  The sides of the cage are set at potential zero to simulate \textquotedblleft boundaries at infinity\textquotedblright.  Since the water vapour tapers off to a constant value sufficiently far away from the crystal, imposing this Dirichlet boundary condition preserves the similarity.  In this analogy, one uses Gauss' flux law and Fick's law of diffusion to establish that $Sh_{Re=0}$ is given by  
\begin{equation}
\label{2}
Sh_{Re=0} = \frac{4\pi CL}{A}
\end{equation}

 where A, L, and C are the surface area, characteristic length, and the crystal's capacitance, respectively.  The units are in cgs for ease of comparison with S\&L78 and Jayaweera(1971).  Thus the problem of diffusion is reduced to finding capacitances of crystals with the Dirichlet boundary conditions mentioned above.

\begin{flushleft}
\textbf{3. FINITE LATTICES AND JACOBI'S METHOD}
\end{flushleft}
 
 To compute (2), a numerical scheme for finding the capacitances for crystals of rectilinear shapes was developed. \textquotedblleft Rectilinear shapes\textquotedblright are those that can be constructed with a finite number of straight edges.  

 A Cartesian grid system is established with a finite number of lattice points within the Faraday cage, the outer most points representing the edges of the cage (box).  Each side of the box is at least 20 times larger than the respective sides of the crystal.  The separation between adjacent lattice points can be different along different axis directions. Each of the lattice points are labelled in terms of positive integer indices (i, j, k); where i, j, and k run along the x-axis, y-axis, and z-axis respectively (Fig. 1).  For simplicity, $\mu$ is introduced, which can be either x, y, or z.  Then $\hat{\mu}$ represents the unit vector along $\mu$-axis. Assuming that the grid points are sufficiently close to one another, the partial derivatives of potential \emph{V} can be approximated by the following set of finite difference equations
\begin{equation}
\label{3}
\frac{\partial V(i,j,k)}{\partial\mu}\equiv\frac{V((i,j,k)+\hat{\mu}) - V(i,j,k)}{\delta\mu}
\end{equation}
 where 
\begin{equation}
\label{4}
A(\mu) \equiv V((i,j,k)+\hat{\mu}) + V((i,j,k)-\hat{\mu}) 
\end{equation}
 then the finite difference approximation of 2nd order partial derivatives of \emph{V} can be written as
\begin{equation}
\label{5}
\frac{\partial^{2}V(i,j,k)}{\partial\mu^{2}}=\frac{A(\mu)-2V(i,j,k)}{2(\delta\mu)^{2}}
\end{equation}
 For convenience, following proportionality constants are introduced:
\begin{equation}
\label{6}
\alpha\equiv(\frac{\delta y}{\delta x})^{2} \ \ \ \ \ \ \ \ \ \ \ \ \ \ \ \beta\equiv(\frac{\delta z}{\delta x})^{2}
\end{equation}
 Then the Laplace's equation in our discrete rectilinear grid system becomes
\begin{equation}
\label{7}
0\approx \frac{B(x)}{2(\delta x)^{2}}+\frac{B(y)}{2\alpha(\delta x)^{2}}+\frac{B(z)}{2\beta({\delta x})^{2}}
\end{equation}
 where B($\mu$)$\equiv$ A($\mu$) - 2V(i,j,k). From (7), one can solve for potential $V(i,j,k)$ at any lattice point (i,j,k) in our Faraday cage.  By rearranging the terms in (7), the following is obtained:
\begin{equation}
\label{8}
V(i,j,k) =  \frac{\alpha\beta A(x) + \beta A(y) + \alpha A(z)}{2(\alpha\beta + \alpha + \beta)}
\end{equation}

 This is the expression used by the \textquotedblleft Jacobi's\textquotedblright iterative scheme for solving
Laplace equation.  An algorithm is constructed that runs with (8).  First, the values of V are assigned to the sides of the Faraday cage and the surface of the crystal, which is sitting in the middle of this Faraday cage.  V is then set to zero on all other lattice points, followed by the running of a computational scheme for V(i,j,k) iteratively for each lattice point in the box at each iteration.  The computation is started out from the crystal surface, then \textquotedblleft branches out towards the edges of the box\textquotedblright.  This procedure is iterated until convergence by computing the difference $\delta V(i,j,k)$ = $|V_{n}(i,j,k) - V_{n-1}(i,j,k)|$, where $V_{n}(i,j,k)$ denotes the value V(i,j,k) at n-th iteration.  The scheme is halted when a desired \textquotedblleft tolerance bound\textquotedblright is reached.  This occurs when following criterion is met
\begin{equation}
\label{9}
\sum_{(i,j,k)}\delta V(i,j,k) < \epsilon
\end{equation}
 where the sum in (9) is carried out for all grid points in the box.  It is well known from literatures in computational physics that the number of iterations 'r' required to reduce the error given by the sum (9) by a factor of $10^{-p}$ is 
\begin{equation}
\label{10}
r \sim O(pN^{3})
\end{equation}

 where N is the number of grid points in a NxNxN cubic regular Cartesian grid system.  At the end of a run of this algorithm, values of all the V's would have been computed for each lattice point up to the desired precision $\epsilon$.

 Some comments are in order at this stage.  Although no regular Cartesian grid system is used for many of the 
crystal models studied here, the order of convergence will be the same as (10).  There are other computational methods to solve Laplace's equation numerically, such as Gauss-Seidel and the \textquotedblleft simultaneous over-relaxation\textquotedblright(SOR) methods.  The SOR method has a convergence rate of $O(N^{2})$ for 3 dimensional Cartesian grid system.  The classical Jacobi method, however, is one of the simplest algorithms that can be used for our study and is relatively free of the complications in implementation compared to the SOR method.  It was thus selected for this study.

\begin{flushleft}
\textbf{4. EXAMPLE: HEXAGONAL PLATE}
\end{flushleft}

 The hexagonal plate was modelled using a non-regular rectangular grid system, with the proportionality factors 
$\alpha$ and $\beta$ not being equal.  Only a finite number of grid points is available for \textquotedblleft drawing\textquotedblright a hexagon to encapsulate the important features of the crystal, namely its vertices. Not all the vertices of the hexagon lie on lattice points if a regular ($\alpha = \beta$) Cartesian grid system is used. Hence a regular Cartesian coordinate system is not used for an optimal design of the crystal on our finite grid. 
 $\alpha$=3 allows for construction of a hexagon in which all its vertices lie on grid points.  $\beta$ can be chosen to be any positive value since the plate is assumed to be thinner than the lattice separation along the z-axis.  Thus $\beta$=1 was selected, meaning that our plate lies on a single xy-planar layer of lattice points, in the middle of the box.  This \textquotedblleft optimal\textquotedblright representation of the hexagonal plate results in the relationship: $N_{d}=\displaystyle\frac{N_{x}}{2}$ where $N_{d}$ and $N_{x}$ represent the number of points used to represent a diagonal segment, and horizontal segment respectively.  In a similar manner, \textquotedblleft optimal\textquotedblright representation was achieved for other crystal shapes as investigated experimentally by S\&L78 in a \textquotedblleft liquid tunnel\textquotedblright (Schuepp and List, 1969).

\begin{flushleft}
\textbf{5. DISCRETE VERSION OF GAUSS' FLUX LAW AND SCALING RULES}
\end{flushleft}

 A simple method of numerically computing $Sh_{0}$ using the potential \emph{V} for lattice points around the crystal involves computation of the capacitance C.  This is done by applying Gauss' law to compute the \textquotedblleft surface charge\textquotedblright of the crystal, via C=$\displaystyle\frac{Q}{V_{0}}$, where $V_{0}$ is the surface potential assigned to the crystal at the onset of Jacobi's method mentioned previously.  The total electric flux of the crystal is obtained by first enclosing the crystal in a \textquotedblleft rectangular cage\textquotedblright.  The sides of this cage are just one lattice point away from the closest side of the crystal. Each side of this 
enclosure consists of its grid elements, which are called \textquotedblleft area element grids\textquotedblright. These grids have the same respective lattice spacing as the grid for the Faraday box.  

 Each of the \textquotedblleft area element\textquotedblright is associated with a set of integers \emph{I}.  Then the discrete approximation of Gauss' flux law can be written as 
\begin{equation}
\label{11}
\displaystyle\oint_{S}\vec{E}\cdot dA\approx\sum_{I}-\frac{\partial V(i,j,k)}{\delta\mu}\cdot A_{i}
\end{equation}
 which is just a sum of flux through each of the area elements.  Note that in (11), the fact that on the surface of a perfect conductor, the electric field is always perpendicular to its surface, was utilized.  The flux through an area element whose outward unit normal is in $\pm\hat{\mu}$ direction is approximated discretely by
\begin{equation}
\label{12}
F_{\pm\mu}=-\frac{V((i,j,k)\pm\hat{\mu}) - V(i,j,k)}{\delta\mu}(\delta\nu)(\delta\gamma)
\end{equation}
 where $\nu$ and $\gamma$ are the other two coordinates not equal to $\mu$.  The main reason for placing our rectangular prism enclosure at most one lattice point away from sides of the crystal is that such a construction allows dealing only with $\frac{\partial V}{\partial\mu}$ rather than with the more complicated $\nabla V$, when computing the flux.  It is worth discussing the consequence of (12) for a specific direction $\mu$. For the flux through a face with its unit normal in $\pm\hat{z}$ direction, the flux through this area element (12) reduces to
\begin{equation}
\label{13}
F_{\pm z}= \delta x\frac{\alpha}{\beta}[-V(i,j,k\pm 1) + V(i,j,k)]
\end{equation}
 The main feature of (13) is that the only unknown value is $\delta x$.  The flux through a face in any of the other directions reduce to expressions of the form (13).  In all of these expressions, the only term with unknown value and dimension is $\delta x$.  For solving (8), it is not necessary to know the physical value of $\delta x$ since all the quantities, except for V, are in dimensionless form as proportionality constants $\alpha$ and $\beta$.  Thus, the numerical evaluation of potential \emph{V} does not depend on the choice of length unit; V can be in whatever unit one desires.  But (13) draws attention to the actual physical value that $\delta x$ represents in the grid system. But note that $Q\sim\delta x$, and so $C\sim\delta x$.  Thus, when computing Sh via (2), it is seen that $Sh\sim \frac{(\delta x)^{2}}{(\delta x)^{2}} = 1$.  In other words, although $\delta x$ is unknown throughout our calculations, all $\delta x$'s cancel out at the end because similarity numbers are involved in our modelling.  Only the relative proportions of shapes and boundary conditions are of importance.  The actual physical length scales are irrelevant in computation of dimensionless numbers, Sh and Nu, and this is well demonstrated in our model.

 The total electric flux through the rectangular prism enclosure is calculated by summing the flux (12) through each of its area elements.  From the net flux, the \textquotedblleft surface charge\textquotedblright of crystal is computed, and thus its capacitance.  The Sherwood number Sh at Re=0 is computed using (2), with $\delta x$ carried throughout all these calculations but cancelling out at this last stage.

\begin{flushleft}
\textbf{6. RESULTS AND DISCUSSION}
\end{flushleft}

 The Sherwood number (Sh) for crystal shapes of interest in a cloud physics environment have been computed for zero convection (Re=0) using the methods outlined above.  The results are listed in Table 1.

\hfill

\begin{tabular}{|c|c|c|c|}
  \hline
  Crystal  & Approximate & Numerical & Difference\\
  Shape & Sh & Sh &  \\
  \hline
  HP & 2.74 & 2.84 & +3.64\% \\
  HC & 2.48 & 2.36 & -4.83\% \\
  BB & 4.02 & 3.88 & -3.48\% \\
  SC & 5.50 & 5.70 & -3.51\% \\
  CC & N/A & 5.67 & N/A \\
  \hline
\end{tabular}
\begin{flushleft}
Table 1: Present numerical calculations of dimensionless mass transfers (Sherwood numbers) of hexagonal plates HP, hexagonal columns HC, broad branched dendrites BB, stellar crystals SC, and capped columns CC; comparisons with previous values of Sh approximated by Mc63(\textquotedblleft Approximate Sh\textquotedblright) and extrapolations by S\&L78, with differences.
\end{flushleft}

 The \textquotedblleft approximate capacitances\textquotedblright which lead to the corresponding approximate Sh were given by Mc63.  Using the analytical calculations of capacitances of spheres and thin circular disks of radius r to approximate the capacitances, Mc63, S\&L78, and Jayaweera have given expressions for capacitance for the shapes listed in Table 1 (with the exception of CC).

\hfill

\begin{tabular}{|c|c|c|}
  \hline
  Crystal & Approximate & Dimensions\\
  & capacitance & used\\
  \hline
  HP & 0.567r & r=8($\delta x$)   \\
  \hline
  HC & $\frac{a}{ln(\frac{2a}{b})}$ & a=12.5($\delta x$) \\
  & & b=8($\delta x$) \\
  \hline
  BB & 0.554r & r=24($\delta x$)  \\
  \hline
  SC & 0.439r & r=6($\delta x$)   \\  
  \hline
\end{tabular} \\
Table 2.: Approximate crystal capacitances (Mc63, S\&L78) [cgs units] and dimensions used for the calculations in Table 1. Note that no analytical or approximate expression of the capacitance of CC were given in either Mc63 or S\&L78.

\hfill

 In summary, the numerical methods developed and applied here provide convincing values of the molecular diffusion of water vapour to ice crystals of various shapes when compared with the values obtained by approximations and extrapolations (Mc63).  It is also of interest to see that the Sherwood number for capped columns is not much different from the value for columns and that the ordering of the values for Re=0 is consistent with the measurements of S\&L78 at higher Reynolds numbers.

\begin{flushleft}
\textbf{ACKNOWLEDGEMENTS}
\end{flushleft}

The authors wish to thank the Natural Sciences and Engineering Research Council (NSERC) of Canada for its support.

\begin{flushleft}
\textbf{REFERENCES}
\end{flushleft}

\begin{flushleft}
K.O.L.F.  Jayaweera, 1971:  Calculations of ice crystal growth.  \emph{\underline{J. Atmos. Sci.}} \textbf{28}, 728-736.

\hfill

J.E.  McDonald, 1963:  Use of electrostatic analogy in studies of ice crystal growth.  \emph{\underline{Z. angew. Math. Phys.}}, \textbf{14}, 610-620.

\hfill

P.H.  Schuepp and R.  List, \textquotedblleft Mass transfer of rough hailstone models in flows of various turbulence levels\textquotedblright, \emph{\underline{J. Appl. Meteor.}}, \textbf{8}, No. 2, 254-263, 1969.

\hfill

R.S.  Schemenauer and R.  List, \textquotedblleft Measurements of the convective mass transfer of planar and columnar ice crystals\textquotedblright, \emph{\underline{Borovikov Memorial Issue}}, Academy of Sciences of the USSR, Leningrad, 217-232, 1978.

\end{flushleft}
 
\end{document}